\documentclass[aps,prb,twocolumn,showpacs,superscriptaddress]{revtex4-1}
\usepackage[latin1]{inputenc}
\usepackage{amsmath}
\usepackage[english]{babel}
\usepackage{graphicx, color}
\usepackage{hyperref}
\definecolor{link}{rgb}{0.1,0.1,0.9}
\hypersetup{colorlinks=true,linkcolor=link,citecolor=link,urlcolor=link,linktocpage}
\usepackage{epstopdf}
\usepackage{color}
\usepackage{amsmath}
\usepackage{longtable}
\usepackage{amssymb}
\usepackage{siunitx}
\usepackage{amsfonts}
\usepackage{float}

\begin{document}
	\title{ Observation of planar Hall effect in Type-II Dirac semimetal PtTe$_{2}$}

 \author{Amit Vashist$^1$, R. K. Singh$^1$, Neha Wadehra$^2$, S. Chakraverty$^2$, and Yogesh Singh}
 \affiliation{Department of Physical Sciences,
 	Indian Institute of Science Education and Research Mohali,
 	Sector 81, S. A. S. Nagar, Manauli, PO: 140306, India.\\
	$^2$ Nanoscale Physics and Laboratory, Institute of Nanoscience and Technology, SAS Nagar, Mohali, Punjab 140062, India.}
	
	\date{\today}

	\begin{abstract}
We report experimental observation of the Planar Hall effect (PHE) in a type-II Dirac semimetal PtTe$_2$. This unusual Hall effect is not expected in nonmagnetc materials such as PtTe$_2$, and has been observed previously mostly in magnetic semiconductors or metals. Remarkably, the PHE in PtTe$_2$ can be observed up to temperatures near room temperature which indicates the robustness of the effect. This is in contrast to the chiral anomaly induced negative longitudnal magnetoresistance (LMR), which can be observed only in the low temperature regime and is sensitive to extrinsic effects, such as current jetting and chemical inhomogeneities in crystals of high mobility. Planar Hall effect on the other hand is a purely intrinsic effect generated by the Berry curvature in Weyl semimetals.  Additionally, the PHE is observed for PtTe$_2$ even though the Dirac node is $\approx 0.8$~eV away from the Fermi level. Thus our results strongly indicate that PHE can be used as a crucial transport diagnostic for topological character even for band structures with Dirac nodes slightly away from the Fermi energy.
		
	\end{abstract}
	
	\maketitle
	
	\section{Introduction}
	
Recently, a new classification scheme of insulators, superconductors, and metals based on quantized topological invariants of the bulk band structure has been constructed \cite{qi2010quantum,RevModPhys.90.015001}. On the basis of this classification, new states of quantum matter such as topological insulators and topological metals have been proposed and discovered very recently \cite{hasan2010colloquium,RevModPhys.90.015001}.  The nontrivial nature of the topological bands is determined by the Berry phase accumulated by the electronic wave function of the charge carrier on traversing an orbit enclosing the crossing point (Dirac point) of gapless linear dispersion in two and three dimensions \cite{hasan2010colloquium,qi2010quantum,RevModPhys.90.015001}. The Berry curvature in the energy-momentum space for these class of topological materials acts like an intrinsic magnetic field, which in turn affects the charge carrier dynamics in an unusual way.

Recently an unconventional version of the Hall effect, known as planar Hall effect (PHE) has been theoretically proposed as a signature of the chiral anomaly in topological semimetals, the origin of which is the large Berry curvature between a pair of Weyl nodes \cite{PhysRevLett.111.246603,PhysRevX.5.031023,PhysRevB.86.115133,Hirschberger2016,Xiong413}. Unlike ordinary Hall effect, the PHE does not require a perpendicular magnetic field, and a transverse voltage is generated by the application of a magnetic field, applied in the plane of the sample/current.  While a small PHE has been reported for ferromagnetic semiconductors and semimetals, PHE has recently been experimentally observed in several classes of magnetic and nonmagnetic Dirac and Weyl semimetals (DSM \& WSM)\cite{1801.05929,PhysRevB.97.201110,PhysRevB.98.161110,PhysRevB.98.041103,li2018giant,PhysRevB.98.081103,gopal2018observation,PhysRevLett.90.107201,PhysRevB.76.035327}. The origin of PHE in ferromagnets has been attributed to several microscopic phenomenon, including spin Hall magnetoresistance, anisotropic scattering by the impurities and non-spherical Fermi surfaces.  On the other hand PHE in Topological materials has its origin in the finite Berry curvature of the nontrivial bands in the bulk. The PHE could be an intrinsic and universal effect like anomalous Hall effect.  Experimental observation of all three effects, chiral anomaly driven negative MR, anomalous Hall effect, and PHE, in the nonmagnetic topological semimetal $ZrTe_5$, strongly suggests a common and intrinsic origin of all these phenomena\cite{Liang2018,li2016chiral,li2018giant}.      
	
Here we report the observation of the Berry curvature induced PHE in the type-II Dirac semimetal PtTe$_2$. The observation of this effect in nonmagnetic PtTe$_2$ suggests that PHE is a distinct and reliable signature of the chiral character in Dirac and Weyl semimetals, as opposed to the observation of the negative longitudinal magnetoresistance (NMR), the origin of which, can be many other extrinsic effects. These effects include primarily current jetting and chemical inhomogeneity in crystals with a very high mobility as is frequently observed in topological semimetals. Therefore identification of the chiral anomaly by negative longitudinal MR is still under debate. \cite{arnold2016negative,dos2016search,Hirschberger2016,Xiong413,Zhang2016}. These experimental complexities, on the other hand, do not affect the PHE and signatures of topological bands have been detected using this effect \cite{1807.06229}. This led us to believe that topological bands could manage to contribute in transport measurements, even in cases when the Dirac or Weyl node is located slightly away from the Fermi energy.  For example, in metallic topological insulators quantum oscillation are observed from surface Dirac states although the chemical potential lies in the bulk conduction/valance bands \cite{Kim2013,PhysRevLett.109.116804,PhysRevLett.109.066803}. 

\begin{figure*}
\includegraphics[width=0.8\linewidth]{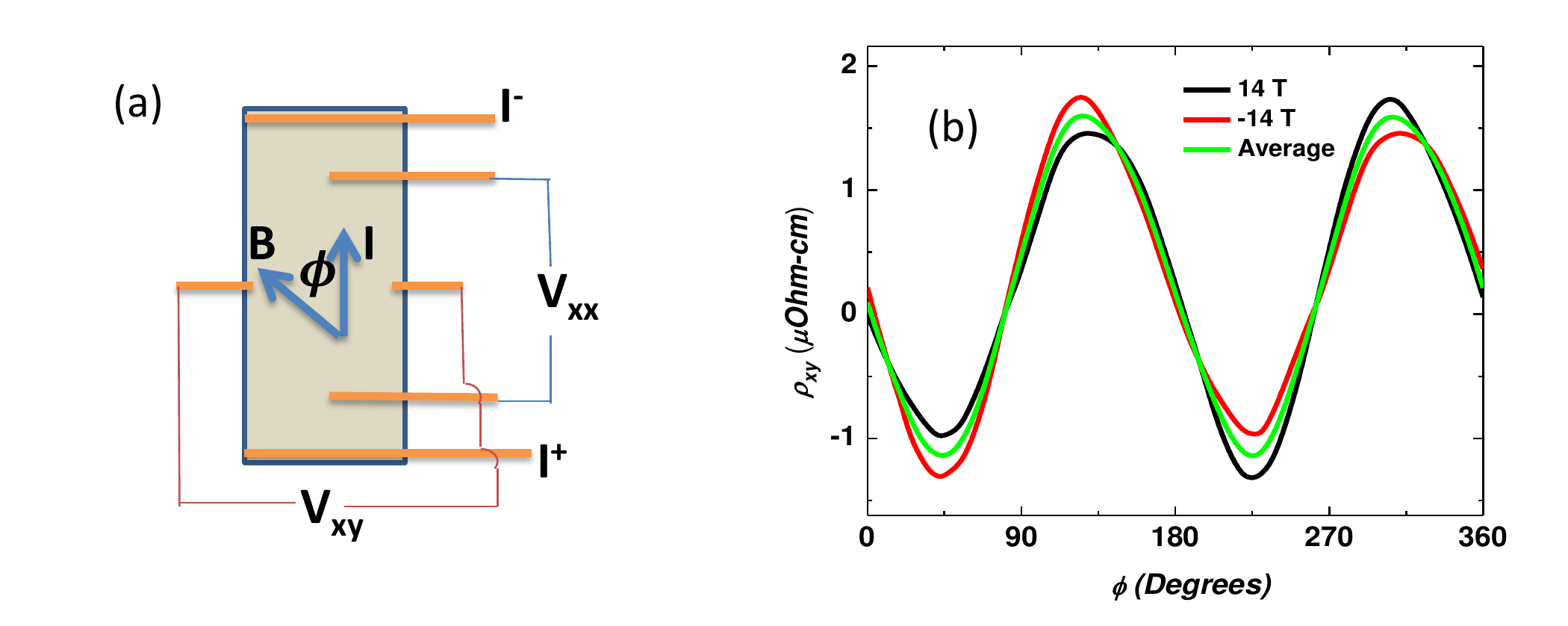}
\caption{(a) Schematic diagram for the planar Hall effect (PHE) measurements. (b) Measured planar Hall resistivity at two opposite field ($14$ and $-14$~T) at $2$~K and their averaged value.}
\label{fig:sch}
\end{figure*}

The magnetotransport properties of the PdTe$_2$ and PtTe$_2$ have been studied in detail recently. These materials are type-II topological semimetals where the three dimensional Dirac cone is tilted towards one of the momentum directions \cite{yan2017lorentz,PhysRevLett.120.156401}. This gives rise to strongly anisotropic transport properties in different crystallographic directions. Owing to the location of Dirac points well below the Fermi level in the PtTe$_2$ (0.8~eV) and PdTe$_2$ (1~eV), the chiral anomaly induced negative MR could not be resolved experimentally, however, the Dirac nature was confirmed by a Berry phase of $\pi$ extracted from transport experiments \cite{PhysRevB.96.041201,PhysRevB.97.235154,pavlosiuk2018galvanomagnetic, PhysRevMaterials.2.114202}. The experimental details including growth of high quality single crystals and transport measurements such as quantum oscillations can be found in our previous reports \cite{PhysRevMaterials.2.114202, PhysRevB.97.054515}. 

In the following we present the observation of PHE in PtTe$_2$ single crystals. The measurement geometry and the field configuration is shown in Fig.~\ref{fig:sch}(a).  The magnetic field $B$ is applied in the plane of the current $I$ and the angle between $B$ and $I$ is controlled by rotating the sample in the plane made by $B$ and $I$ as shown in Fig.~~\ref{fig:sch}(a).  The planar Hall resistivity ($\rho^{PH}_{xy} = V_{xy}/I$) and the planar anisotropic MR ($\rho_{xx} = V_{xx}/I$) are measured simultaneously.  Contributions from the normal Hall effect can arise due to a small misalignment of the sample plane with respect to the magnetic field.   The measurements of the two resistivities is performed in both negative and positive magnetic field polarities and is subsequently averaged and thus any stray contribution from the normal Hall resistivity is removed.  A typical raw data in negative and positive magnetic field polarity and the averaged data as a function of the in plane angle $\phi$ (between the coplanar magnetic field and the current) at $14$~T is shown in Fig.~\ref{fig:sch}(b). 

\begin{figure*}
\includegraphics[width=0.8\linewidth]{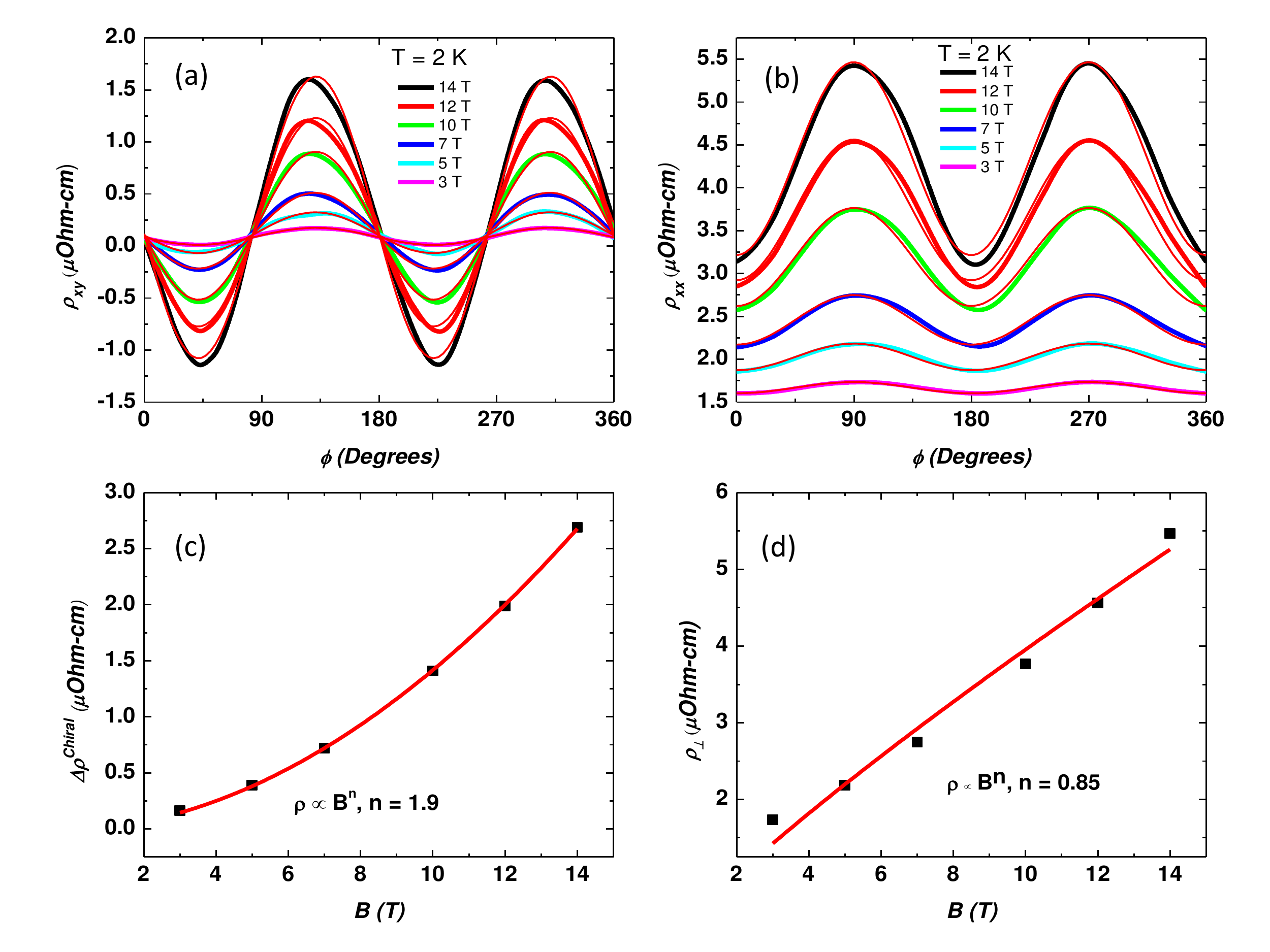}
\caption{(a) \& (b) In plane angular dependence of symmeterized planar hall resistivity ($\rho^{PH}_{xy}$) and longitudinal resistivity ($\rho_{xx}$) in different magnetic fields at $T = 2$~K\@.  The solid curves through the data are fits by equations~(1) and (2). (c) and (d) Extracted values of $\triangle$$\rho$$_{chiral}$ and $\rho$$_{\perp}$ versus magnetic field. The solid curves through the data are power law fits.}
\label{fig:res}
\end{figure*}

The field dependent $\rho^{PH}_{xy}$ and $\rho_{xx}$ at $T = 2$~K are shown in Figs.~\ref{fig:res}~(a) and (b).  It is clear from the data in Fig~\ref{fig:res}(a) that as expected the planar Hall contribution $\rho^{PH}_{xy}$ has distinct minima and maxima at the in plane angular positions $\phi = 45$~\textdegree and $135$~\textdegree, respectively.  On the other hand the variation of the anisotropic MR $\rho_{xx}$ as a function of $\phi$, shows characteristic maxima and minima at $90$~\textdegree and $180$~\textdegree~ as expected theoretically.  A phase difference of $45$~\textdegree~ between the $\rho^{PH}_{xy} - \phi$ and $\rho_{xx} - \phi$ curves is consistent with theoretical predictions. This observation is consistent with previously observed PHE data on other Dirac and Weyl semimetals, such as ZrTe$_5$, GdPtBi, Cd$_3$As$_2$, VAl$_3$, MoTe$_2$ and Co$_3$Sn$_2$S$_2$ \cite{1801.05929,PhysRevB.97.201110,PhysRevB.98.161110,PhysRevB.98.041103,li2018giant,PhysRevB.98.081103,gopal2018observation,PhysRevLett.90.107201,PhysRevB.76.035327}. The value of $\rho^{PH}_{xy}$ increases with increasing magnetic fields up to 14 T, and reaches values of $1.61~\mu \Omega$-cm. This value is quite large in comparison with the values estimated in ferromagnets, but is comparable to values found in recent experiments on some of DSMs and WSMs \cite{1801.05929,gopal2018observation}. However, this value is smaller than those found for GdPtBi and ZrTe$_5$ \cite{PhysRevB.98.041103,li2018giant}, which could be due to the fact that the chemical potential lies near to the Dirac point in these topological semimetals. On the other hand Dirac point in PtTe$_2$ is located $0.8$~eV below the Fermi level.

\begin{figure*}
\includegraphics[width=0.8\linewidth]{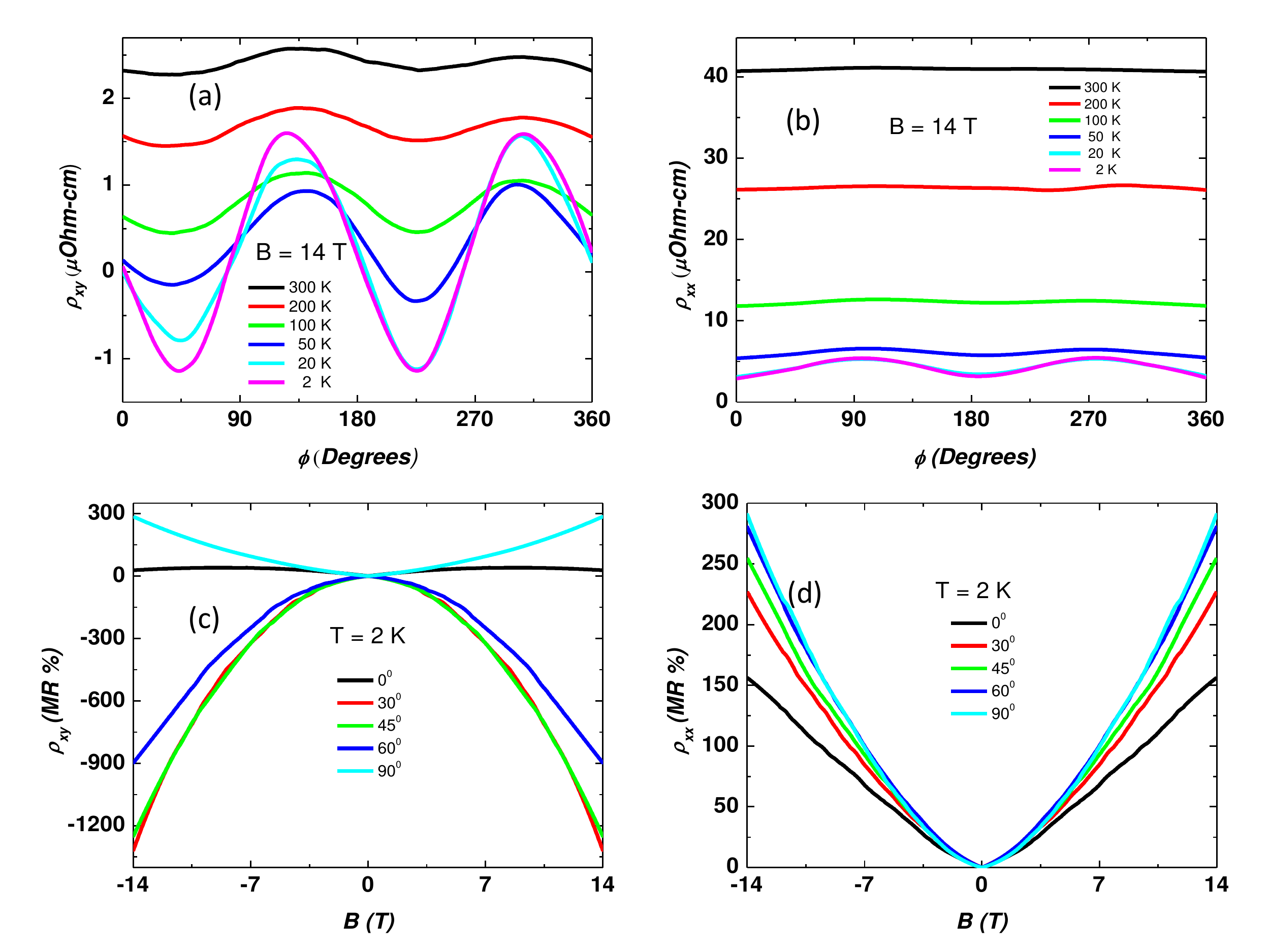}
\caption{(a) \& (b) Angular dependence of unsymmeterized planar hall resistivity ($\rho^{PH}_{xy}$)and longitudinal resistivity($\rho_{xx}$) at different temperature at magnetic field 14 T. (c) \& (d)The symmeterized $\rho^{PH}_{xy}$ and $\rho_{xx}$ at 2 K. }
\label{fig:mr}
\end{figure*}
	
In order to estimate the chiral anomaly contribution $\triangle$$\rho$$_{chiral}$ in the measured $\phi$ dependent $\rho^{PH}_{xy}$ and $\rho_{xx}$ data , we fit the data with the theoretically derived semiclassical expressions of the two resistivities which are given as : \cite{PhysRevLett.119.176804}

\begin{equation}
\rho_{xy}^{PH} = -\triangle\rho^{chiral}\sin\phi\cos\phi+b\cos^{2}\phi+c
\end{equation}  
	
\begin{equation}
\rho_{xx} = \rho_{\perp}- \triangle\rho^{chiral}\cos^{2}\phi
\end{equation}  

The first term in equation~$(1)$ represents the contribution to $\rho_{xy}^{PH}$ from the chiral anomaly.  The $\triangle$$\rho$$_{chiral}$= $\rho$$_{\perp}$-$\rho$$_{\parallel}$, with $\rho$$_{\perp}$ and $\rho$$_{\parallel}$ being the resistivities in the in-plane transverse ($\phi = 90^o$) and parallel ($\phi = 0$) magnetic fields, respectively. The second term in the equation (1) takes into account the possible contribution of the anisotropic MR which might arise from misalignment effects in the measurements, and the third term $c$ accounts for the constant Hall resistivity resulting from any nonuniform thickness of the sample \cite{li2018giant}.  The fitting by the above equations to the $\rho^{PH}_{xy}(\phi)$ and $\rho_{xx}(\phi)$ curves at different fields is shown as the solid curves through the data in Figs.~\ref{fig:res}~(a) and (b).  The extracted value of the $\triangle$$\rho$$_{chiral}$ and the transverse resistivity $\rho$$_{\perp}$, from the fits is plotted as a function of the magnetic field in Figs.~\ref{fig:res}~(c) and (d), respectively.  As expected, the value of $\triangle$$\rho$$_{chiral}$ and $\rho$$_{\perp}$ increases with increasing magnetic field. The value of $\triangle$$\rho$$_{chiral}$ at $2$~K in $14$~T is found to be $\approx 2.7~\mu \Omega$-cm. To quantify the variation of $\triangle$$\rho$$_{chiral}$ and $\rho$$_{\perp}$ with magnetic field we have fitted these data by power-law dependence. On fitting the $\triangle$$\rho$$_{chiral}$ data with the power law function ($\triangle$$\rho$$_{chiral}$ = $B^{n}$)\cite{,PhysRevB.98.081103}, we find the value of $n = 1.9(2)$ which is close to a quadratic field dependence. This is consistent with theoretical expectations.  Similar values of the power law coefficient $n$ has been found in many other DSM and WSM. On the other hand the value of $n$ in the case of $\rho$$_{\perp}$ vs magnetic field is not near to $n = 2$, and varies linearly. The possible reason for this variation are not known yet, and similar dependence has been observed in some of the other WSMs.        

We next investigate the robustness and the intrinsic nature of the PHE, by taking the temperature dependent $\rho^{PH}_{xy} - \phi$  and $\rho_{xx} - \phi$ data at maximum applied magnetic field of $14$~T\@. It is clear from the data shown in Fig.~\ref{fig:mr}~(a) that the PHE survives up to $T = 300$~K, whereas the magnitude of $\rho_{xx}$ - $\phi$ decreases rapidly with $T$ as can be seen from Fig.~\ref{fig:mr}~(b). The periodic nature of the $\rho^{PH}_{xy} - \phi$ curves is maintained up to room temperature and is not destroyed by the electron-phonon scattering. This is due to the fact that the topological and the chiral nature of the Dirac/Weyl electrons makes them immune to backscattering from defects and other electron-electron and electron-phonon scattering mechanisms. This is also why such topological systems show very large mobility and MR. The observation of PHE up to high temperature thus confirms the intrinsic nature of this effect which is driven by the large Berry curvature between the Weyl nodes. Recent observation of the chiral anomaly induced negative MR, anomalous Hall effect and PHE in GdPtBi, Co3Sn2S2 and nonmagnetic $ZrTe_5$, strongly suggest that all three phenomena have a common, intrinsic origin. Moreover, it is a direct indication that the PHE does not require the ferromagnetic nature of the sample and previously observed PHE in semiconducting and metallic ferromagnets might have the origin from the Berry curvature effect.   

Figure~\ref{fig:mr}~(c) and (d) show the magnetic field dependence of the two in plane resistivities measured at various fixed angles $\phi$ and at temperature of 2 K. On varying the angle from $0$~\textdegree to $90$~\textdegree, the magnitude of the $\rho^{PH}_{xy}$ first increases and then decreases, which is a hallmark signature of the PHE as expected theoretically. The in plane anisotropic nature of the $\rho_{xx}$ is evident from the field dependent measurements at various in plane $\phi$, as shown in Fig.~\ref{fig:mr}~(d). On the other hand, we could not resolve the negative MR in the coplanar and collinear current and magnetic fields configuration. This can be seen from the Fig.~\ref{fig:mr}~(d).    
	
In conclusion, we report observation of the planar Hall effect in single crystals of the nonmagnetic Type-2 Dirac semimetal PtTe$_2$. The PHE is observable up to room temperature suggesting the robustness of the relativistic carriers against the electron -phonon scattering. The Dirac/Weyl node in PtTe$_2$ is known to lie about $0.8$~eV below the Fermi level.  We note that recently PHE has also been reported in the iso-structural material PdTe$_2$ which is also a Type-2 Weyl semi-metal.  The Weyl node in PdTe$_2$ is also situated away from the Fermi energy.  Thus our observation of PHE for a system like PtTe$_2$ therefore suggests that PHE can be used as a crucial transport diagnostic for topological character even for band structures with Dirac nodes slightly away from the Fermi energy.

\bibliographystyle{apsrev4-1}
\bibliography{Ref}

\end{document}